\documentclass[a4paper]{jpconf}
\usepackage{graphicx}
\begin{document}
\title{High energy physics from high performance computing}

\author{T. Blum}
\address{Physics Department, University of Connecticut, Storrs, CT
    06269-3046, USA}
    \ead{tblum@phys.uconn.edu}


\begin{abstract}
We discuss Quantum Chromodynamics calculations using the lattice regulator. The theory of the strong force is a cornerstone of the Standard Model of particle physics. We present USQCD collaboration results obtained on Argonne National Lab's Intrepid supercomputer that deepen our understanding of these fundamental theories of Nature and provide critical support to frontier particle physics experiments and phenomenology. 
\end{abstract}

\section{Introduction} Thanks to recent advances in computer power, software, and algorithms, lattice gauge theory calculations of Quantum Chromodynamics (QCD), or Lattice QCD for short, have come a long way since Wilson proposed the method\cite{Wilson:1974sk} 35 years ago. In this talk, we discuss results from lattice QCD and members of the USQCD collaboration that impact frontier particle physics research.

QCD is the fundamental theory of the strong force responsible for the material world. Quarks and gluons, interacting via QCD bind together to form hadrons like the ubiquitous protons and neutrons and less well known,  less massive $\pi$ and K mesons, in some ways like electrons and nuclei form atoms by exchanging the electromagnetic force carriers, the photons. QCD is different from Quantum Electrodynamics (QED) in some important ways. For one, the coupling constant, or strength of the QCD interaction, is large, unlike the electromagnetic charge $e$. Further, QCD is said to confine: quarks and gluons which carry the ``color" charge of QCD are not free; they only exist in bound states that are neutral with respect to color charge. These eigenstates of the theory are the hadrons just mentioned.  Finally, there are eight force carriers in QCD, called gluons, compared to one photon in QED, and they interact with each other as well as the quarks. For these reasons, QCD at the relatively low energy scale of a typical hadron mass, say 1 GeV,  can not be studied in a perturbative framework. Being a relativistic quantum field theory, QCD must be regularized, or rigorously defined. This can be done using a discrete hypercubic lattice of points to describe space-time which we come back to in the next section.  

Discretizing the continuum theory leads naturally to numerical simulations. The non-abelian nature of QCD and the fact that an almost exact chiral symmetry of the quarks exists in Nature make it difficult to solve, even with the largest supercomputers. In this talk we describe the state-of-the-art simulations being done on some of the largest supercomputers in the world, the algorithms used for them, and the theoretical physics results gained from them.

\section{Dynamical fermion simulations}
The heart of any lattice gauge theory (see, {\it e.g.}, \cite{Creutz:1984mg}) calculation is its ensemble of gauge configurations, representing the gauge
fields to be averaged over in the path integral for the quantum expectation value of some observable. For lattice QCD these configurations of ``link" variables represent the gluons, for QED the photons. Feynman's path integral is to be taken over all possible configurations, or paths of the system, in configuration space. Since a configuration is the specification of the fields at every point in space-time, and since the (components of) fields take all possible values between $\pm\infty$, the un-regulated path integral diverges. It is made finite by discretizing the continuum Lagrangian
to produce one that exists on a four dimensional hyper-cubic lattice with spacing $a$. The continuum limit, $a\to0$, is taken practically by computing at several values of $a$ and then extrapolating to 0. In practice the 4-volume is finite too, $VT=(N_s a)^3\times N_t a$ ($N_{s,t}$ are the number of lattice sites in the space and time directions), so the infinite volume limit is taken in a similar way.

There are several (quark) degrees of freedom at each lattice site and several (gluon) degrees of freedom on each link connecting neighboring sites in a typical simulation, and $N_s\sim 16-64$, $N_t\sim 32-200$,
so enumerating {\it all} possible configurations is impossible. We are therefore lead to Monte-Carlo simulations using importance sampling to estimate the path integral\cite{Toussaint:1988ft}. The configurations are chosen randomly, but with a weight equal to the exponential of their action. A typical configuration is generated from a Markov chain of configurations by performing an ``update", or change, on the previous configuration. As long as the update algorithm satisfies basic constraints like detailed balance and ergodicity, in the limit of many updates, one is guaranteed to generate configurations with the correct weight, even if the starting configuration is not part of the equilibrium distribution~\cite{Toussaint:1988ft}. Of course, there are many update algorithms that satisfy the basic requirements, so the trick is to find one that is optimal, that is, one that samples the configuration space as efficiently, but as cheaply, as possible. The USQCD collaboration has been using  a combination of  the hybrid monte-carlo\cite{Gottlieb:1987mq,Duane:1987de} (HMC) and rational HMC\cite{Clark:2006fx} algorithms. The former is used for an even number of degenerate fermions, while the latter must be used for an odd number. The RHMC can also be used for even numbers of fermions. 
Here hybrid refers to the two steps in the update algorithm. First the fields are evolved through phase space by adding a fictitious conjugate momentum for the gauge field and numerically integrating the coupled Hamilton's equations of motion for the system in ``monte-carlo time". The trajectory length is defined as the number of integration steps times $\Delta t$, the integration time-step. Typical trajectory lengths have $O(1)$ time-units. This first step is often referred to as the molecular dynamics update. Since the integration is inexact (an error of order $\Delta t^2$ accrues in the hamiltonian after each trajectory), a second metropolis accept/reject step is performed after each trajectory which makes the algorithm exact, {\it i.e.,} the distribution of gauge configurations is the correct one in the limit that the number of updates is large.

State-of-the-art calculations employ 2+1 flavors of {\it dynamical} light-mass quarks, meaning two degenerate up (u) and down (d) quarks, and one heavier strange quark (s). In Nature, the u and d quarks are nearly massless ($(m_u+m_d)/2\sim 3-5$ MeV), while the s quark is somewhat heavier ($m_s\sim 90-100$ MeV).  Dynamical means the virtual pair creation/annihilation of the quarks, or vacuum polarization, effects on the gauge fields are included in the gauge configuration generation. In correlation functions, or expectation values, valence quarks appear, including heavier flavors of quarks. In Nature there are three, charm (c), bottom (b), and top (t). These quarks are so heavy ($\sim 1-175$ GeV) that their vacuum polarization effects are ignored compared to the dominant statistical and systematic uncertainties in present lattice calculations. However, lattice results are nearing the one-percent precision level for some quantities, so groups are starting to consider including the lightest of the heavies, the c quark, in future configuration generation. Of the heavies, only charm can be simulated directly on present day lattices ($a>1/3$ GeV$^{-1}$) because the dimensionless mass should satisfy $m a\ll1$ 
to avoid lattice spacing artifacts. There are ways to treat the heavy quarks so that even b quark observables may be computed accurately. Some of these are described later. The t quark is so massive that its decays can be treated accurately in perturbation theory.
\begin{center}
\begin{table}[h]
\begin{minipage}{16pc}
\caption{\label{tab:asq confs}Asqtad gauge configurations generated in 2008 by the MILC collaboration on the ANL BG/P (Intrepid). ``spc" is the lattice spacing in fm, and ``traj" is the number of trajectories for the entire Markov chain.}
\centering
\begin{tabular}{@{}*{7}{l}}
\br
spc   &  $m_l/m_s$    & $m_\pi L$    & size & traj\\
\mr
\verb 0.06 &                       0.10   &        4.3 &         $64^3\times 144$ & 4956\\
\verb 0.06     &                   0.15       &     4.4    &      $56^3\times144$& 4800 \\
\verb 0.045     &                 0.20         &   4.6      &    $64^3\times192$& 5166 \\
\verb 0.09          &              0.05            &4.8         & $64^3\times96$&4500\\
\br
\end{tabular}
\end{minipage}
\hspace{3pc}
\begin{minipage}{14pc}
\caption{\label{tab:dwf confs} Same as Table~\ref{tab:asq confs}, but for (5d) DWF, generated by LHPC, RBC, and UKQCD collaborations.}
\begin{tabular}{@{}*{7}{l}}
\br
spc.   &  $m_l/m_s$    & $m_\pi L$    & size & traj\\
\mr
\verb 0.08 &                       0.15   &        4.1 &         $32^3\times 64\times 16$ & 6000\\
\verb 0.08     &                   0.22       &     4.9    &      $32^3\times64\times 16$ & 6000\\
\verb 0.08     &                 0.29        &   5.5      &    $32^3\times64 \times 16$& 6000\\
\verb 0.08     &                 0.11        &   4.6      &    $48^3\times64 \times 16$& 400\\
\br
\end{tabular}
\end{minipage}
\end{table}
\end{center}
\vskip -2pc

The quarks (fermions) always appear as the inverse of a large, sparse matrix known as the propagator because it describes how the quarks move through space-time. The matrix ($D$) is the finite lattice representation of the so-called Dirac operator which appears in the quantum mechanical equation of motion for a relativistic spin 1/2 particle. In the continuum the Dirac operator is proportional to a first order 4d covariant partial derivative. They only enter through $D$ because the quantum fields themselves are anti-comuting, complex
numbers known as Grassman variables which we do not know how to simulate on a computer. Since they appear quadratically in the action, sandwiching the Dirac operator, their contribution to the path integral can be computed analytically,
leaving only factors of the propagator and a square root of the determinant of $D$ for each flavor of quark in the action (see~\cite{Peskin:1995ev}, for example). Evaluating the factors involving $D$ dominates any calculation with fermions and boils down to inverting $D$, even for the determinant factors. This is because the $\det D$ factors can be re-exponentiated to the action by introducing fictitious bosons called pseudofermions. Their presence is felt through a force term in the equations of motion during the molecular dynamics update\cite{Gottlieb:1987mq}. The cost due to the gauge part of the action is almost negligible by comparison, and most correlation functions of interest involve fermions.

There is much freedom in defining $D$; it is only required to equal the continuum operator in the limit $a\to0$ and should also be gauge invariant to avoid unwanted mixing with unphysical operators. Thus, several discretizations are employed which are equivalent up to small errors proportional to positve powers of $a$ (usually two). Within USQCD three types of fermion operators are being used in gauge field generation: $a^2$-tadpole improved (Asqtad), domain wall and $a^2$ improved Wilson, or ``clover" fermions. All three formally differ from the continuum theory by order  $a^2$ errors, and they each have strengths and weaknesses. We discuss the first two, the latter is discussed in the talk by Edwards. 

\subsection{Asqtad fermions}
Asqtad fermions are  improved staggered fermions\cite{Lepage:1998vj,Orginos:1998ue,Orginos:1999cr} which ameliorate the flavor symmetry breaking $O(a^2)$ errors of un-improved staggered fermions. Staggered fermions reduce the number of fermion degrees of freedom due to lattice fermion doubling\cite{Nielsen:1981hk}
by diagonalizing the naive discretization of the Dirac operator and then eliminating 3 out of 4 of the spin components on each lattice site. In the limit $a\to 0$, the remaining spin degrees of freedom correspond to four degenerate Dirac fermions. The weight for a single fermion is then given by $(\det D)^{1/4}$.  If $a\neq0$, the flavor symmetry (called taste-symmetry to distinguish it from a continuum flavor symmetry) is broken. It is this symmetry that is partially restored by the Asqtad formulation. There has been vigorous debate in the literature over the validity of  the ``rooting" procedure (whether it yields a continuum Dirac fermion, or a sick cousin when $a\to0$)\cite{Creutz:2007yg,Bernard:2007eh,Bernard:2007ma,Creutz:2008kb}).  The primary advantages of staggered fermions are that they are relatively cheap to simulate and they have an exact axial symmetry for $a\neq0$ left over from the full $SU(3)_L\times SU(3)_R$ chiral symmetry of continuum QCD.

\begin{figure}[h]
\begin{minipage}{16pc}
\hskip 0pc 
\includegraphics[width=15pc]{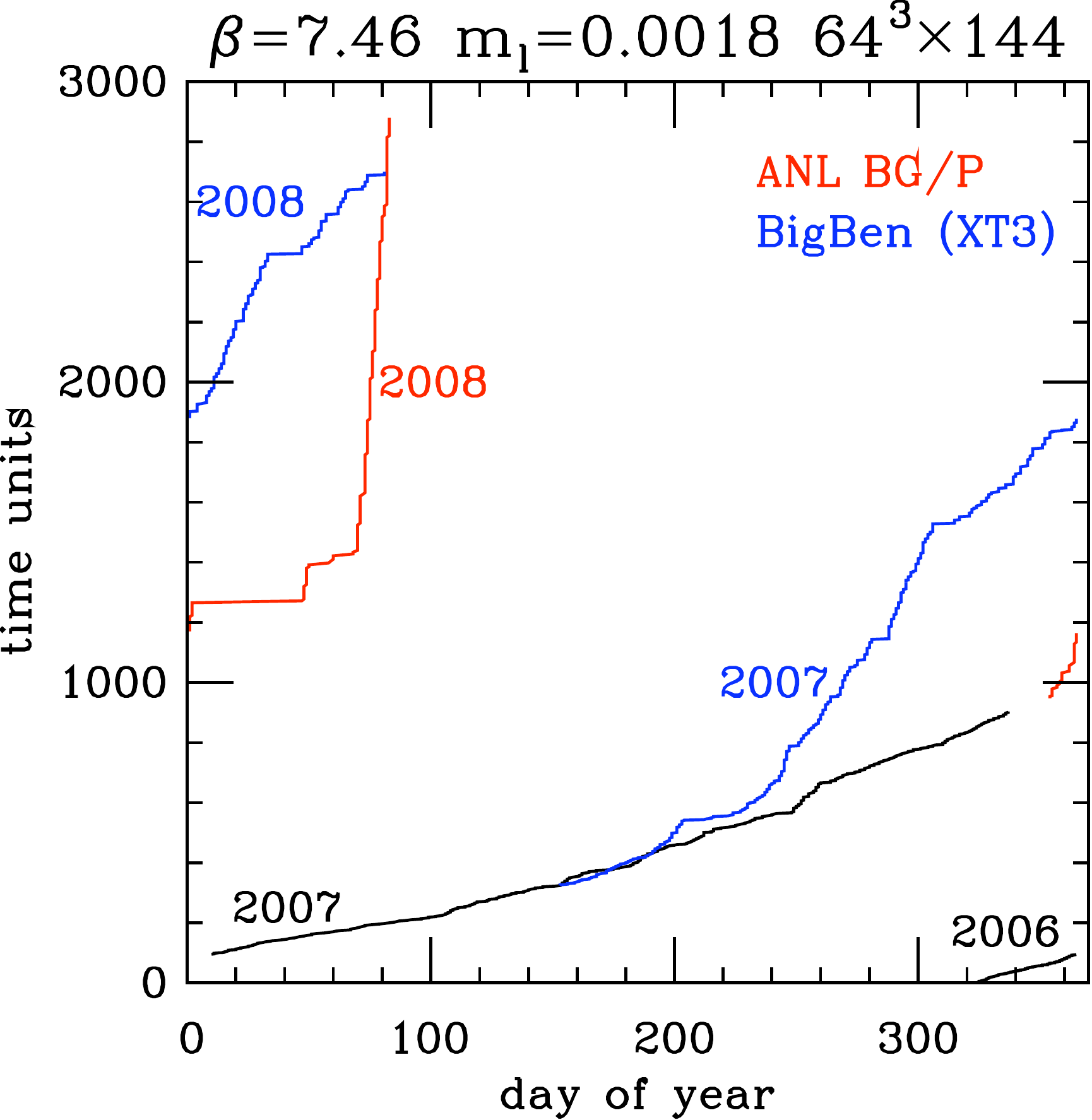}
\caption{\label{fig:asq time}Time history of Asqtad fermion gauge configuration generation. The ensemble is described in the first row of Table~\ref{tab:asq confs} and corresponds to the lightest, most difficult quark mass.}
\end{minipage}\hspace{1.pc}%
\begin{minipage}{21pc}
\hskip 0pc 
\includegraphics[width=21pc]{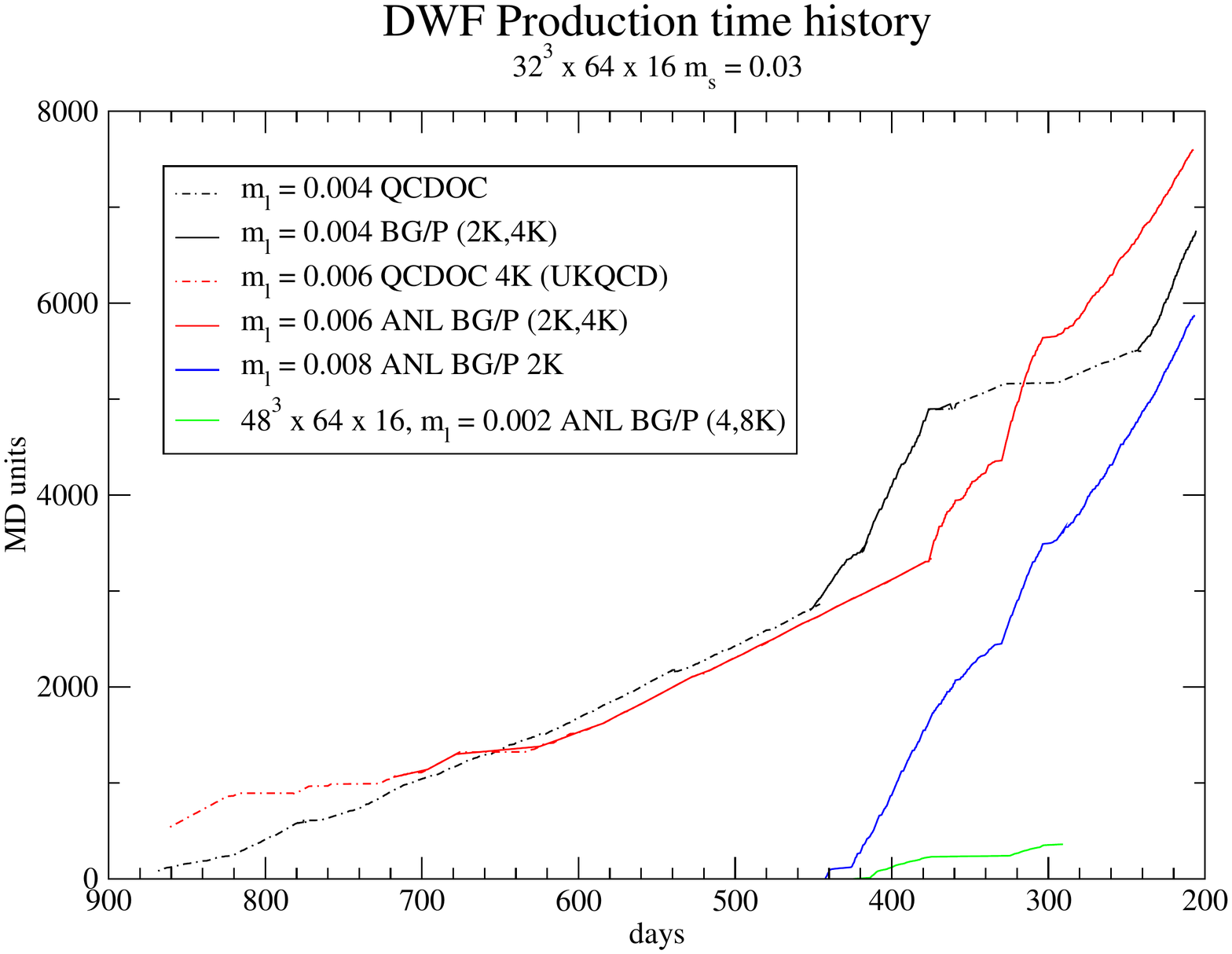}
\caption{\label{fig:dwf time}Time history of DWF fermion gauge configuration generation. The ensembles are described in Table~\ref{tab:dwf confs}.}
\end{minipage}\hspace{2pc}%
\end{figure}

\subsection{Domain wall fermions}

Domain wall fermions (DWF) were first formulated in~\cite{Kaplan:1992bt} as a non-perturbative method for defining and simulating chiral gauge theories,  but were quickly reformulated for numerical efficiency~\cite{Shamir:1993zy}. In~\cite{Shamir:1993zy} it was also recognized that DWF could be used to simulate vector gauge theories like QCD with exact chiral symmetry even when $a\neq0$.  This is the principle advantage of DWF: chiral symmetry protects the theory from unwanted lattice artifacts and renders the lattice theory continuum-like even when $a\neq0$. The symmetry only becomes exact when the size of an extra 5th dimension, which provides the chiral symmetry to low-energy, 4 dimensional fermions stuck to the boundaries of the extra dimension, is taken to $\infty$. For finite size the chiral symmetry is softly broken, inducing a small shift in the quark mass in addition to the bare mass appearing in the lagrangian. This shift, however, can be computed accurately from the simulation and its effects eliminated, or subsumed into the quark mass. The addition of an extra dimension, of course, increases the computational burden. The expectation is that the extra burden is partially offset by the continuum-like character of DWF since $a$ may be relatively larger when continuum symmetries are intact. 

\subsection{Early Science Program at ALCF}
In 2008, the USQCD collaboration was awarded Early Science Program (ESP) time on the Blue Gene/P Intrepid supercomputer at the Argonne (National Lab) Leadership Computing Facility (ALCF) under our three year INCITE allocation. The ESP time was primarily used for gauge field ensemble generation. The stability of Intrepid combined with USQCD's readiness to compute efficiently at the start of operation lead to spectacular gains in ensemble generation, advancing our science goals approximately two years ahead of schedule. The configurations are summarized in Tables~\ref{tab:asq confs} and ~\ref{tab:dwf confs} and the generation dramatized in Figs.~\ref{fig:asq time} and~\ref{fig:dwf time}. 

In the chiral limit of the light quarks, $m_l\to0$, $D$ becomes singular and can not be inverted. Even though the physical u and d quarks are most likely massive, their masses are sufficiently small that present simulations are run with heavier masses, and expectation values are obtained by extrapolating to the physical point. Even though calculations could be done for lighter masses, even physical ones, the volumes would be too small to accommodate such light particles. A rule of thumb for QCD is that the mass of the pion ($m_\pi$) times the linear size of the box should be greater than 4 to avoid significant finite volume effects (see Tables~\ref{tab:asq confs} and \ref{tab:dwf confs}). Recent work suggests this may be an underestimate of the needed size for some quantities~\cite{Yamazaki:2008py}.

Besides several values of the light quark mass for each type, the various ensembles of configurations summarized in Tables~\ref{tab:asq confs} and~\ref{tab:dwf confs} 
reflect the exigencies of the two discretizations. The Asqtad configurations cover three lattice spacings (known as fine, super fine, and ultra fine) to bring the taste-breaking artifacts under control and together with existing coarser lattice spacing ensembles to take the continuum limit. The values of $a$ and $m_l$  are such that very large lattice sizes are required to simulate sufficiently large physical volume.  For the DWF configurations, the lattice spacing is not as small, but is significantly smaller than a companion ensemble with $a=0.114$ fm which together with these is being used to take the continuum limit\cite{Allton:2008pn} (recall the leading artifacts for both go like $a^2$). The size of the extra dimension is 16 lattice sites for each. The last row in Table~\ref{tab:dwf confs} refers to a very demanding large volume, small mass simulation that was begun on Intrepid. However, the added size of the lattice and decreased mass have pushed the update algorithm beyond present capabilities, despite significant gains from tuning which we turn to next.  Completion must await the next generation of leadership class computers.

\subsection{Tuning the rational hybrid monte carlo for domain wall fermions}
Improvements in algorithms have played a large role in lattice QCD
simulations in the last few years.  Most of the time used in
generating lattices is related to including the full effects of
light quarks in the Feynman path integral.  The quarks enter through
the determinant of the Dirac operator, which is a function of all
the gluon fields on the lattice, and which also prohibits two quarks
from occupying the same state, as required by the Pauli exclusion
principle.  The determinant is a positive definite quantity,
so it can be added to an importance sampling measure. The standard
way to do this, by writing the determinant as a bosonic path integral,
requires all eigenvalues be positive.  This is not the case here.
However, the eigenvalues come in related pairs, which taken together are
positive.

To handle a single quark correctly, the determinant of the product
$D \, D^\dagger$ is used,
which is manifestly positive eigenvalue-by-eigenvalue.  On can show that
$\det(D \, D^\dagger) = \det(D) \det(D)$, so that the simulation of a single
quark requires that $(\det(D \, D^\dagger)^{1/2}$ be known.  Using a
rational approximation for $(\det(D \, D^\dagger)^{1/2}$ yields an accurate
way to include single quarks into simulations, which is vital for the
2+1 flavor simulations that most accurately represent the physical world,
where the up and down quarks are taken as light and degenerate in mass,
and the strange quark is much heavier \cite{Clark:2006fx}.

The RHMC algorithm,
as currently implemented, uses a rational approximation for the
effects of a single quark species, along with many other improvements,
to allow the molecular dynamics steps to be as large as possible.
We would like to detail a few of the recent improvements used in the
RHMC algorithm.

In every step of the molecular dynamics part of the RHMC algorithm, the
propagation of quarks in the gluon background must be determined.  This is
the most numerically expensive part of the calculation and is generally
done by a Krylov space solver.  As quark masses get smaller (the physically
interesting region) the solutions are ever more expensive.  One feature
added to current RHMC implementations is to minimize the force from light
quarks via the insertion of Hasenbush preconditioning masses
\cite{Hasenbusch:2001ne}.  The light quark forces are still expensive to evaluate,
but they are not evaluated as often, since they are small.  This allows
one to use an integrator with multiple time scales:  expensive to calculate
forces are made small and only evaluated on a relatively large time scale,
while large, cheaper forces are evaluated often.  Use of multiple time scale
integrators has allowed us to target our computing time most effectively.

The improvements mentioned above have been in place for the last
2-3 years.  For the quark masses and voluems we were simulating
with 2-3 years ago, speed-ups of a factor of 5 or more were seen.
This was a sufficiently large gain that we then moved to smaller
quark masses which were not possible without the improved RHMC
algorithm.  In the last year, with the availability of the large
BG/L and BG/P installations, we have been doing simulations at much
smaller quark masses, and on much larger volumes.  In our our most
demanding simulation to date (last row, Table~\ref{tab:dwf confs}), we observed that it was taking
over 3 times as long as expected.  This DWF simulation has a pion mass of roughly 240 MeV which
is small by current standards, but still well above
the physical value of 140 MeV.

It is important to know whether we had reached a fundamental limit in
the capability of the RHMC, related to physics
at such a light quark mass, or whether the particular choice of RHMC
parameters was not optimal.  Fortunately, we have found the latter to
be the case.  For this light quark mass, we introduced an additional
Hasenbush preconditioning mass and an additional time step in our
multi-time step integrator.  These steps should have helped to keep the
forces from the light quarks small and allowed us to use a large time
step for them.  Initially these improvements did not help with the
acceptance.  However, we then decreased the stopping condition for our
Krylov solvers by a factor of 100, to $3 \times 10^{-8}$.
With this increased accuracy, we were able to increase the molecular dynamics step size
and regain the factor of three we had lost for this simulation.
Our system has $O(10^9)$ degrees of freedom and, for a good acceptance,
the change in the energy at the end of a molecular dynamics trajectory must
be an $O(1)$ quantity.  Thus it is not surprising that such a small residual
is required in our Krylov solvers.

For our current 2+1 flavor simulations, we have found it optimal to rewrite the path integral as
\begin{equation}
  Z  =  \int \, [dA] \,
  \left( \frac{\det {\cal D}(m_l)}
  {\det {\cal D}(m_x)} \right)
  \left( \frac{\det {\cal D}(m_x)}
  {\det {\cal D}(m_s)} \right)
  \left[
  \left( \frac{\det {\cal D}(m_s)}
  {\det {\cal D}(1)]}
   \right)^{1/2} \right]^3
  \exp \left\{ - S_g (A) \right\} 
  \label{eq:zqcd_det_nf2+1}
\end{equation}
where ${\cal D}(m) = D^\dagger(m_l)  D(m_l)$
and $m_l = m_u = m_d$.  Each ratio of determinants involves a separate
pseudofermion field and the fermions are integrated on three different
time scales. $A$ represents the gluon field, and $S_g$ the action of its kinetic term.

\section{Results}\label{sec:results}

\subsection{Improved staggered fermions}

\begin{figure}[h]
\begin{minipage}{18pc}
\hskip 0pc 
\includegraphics[width=18pc]{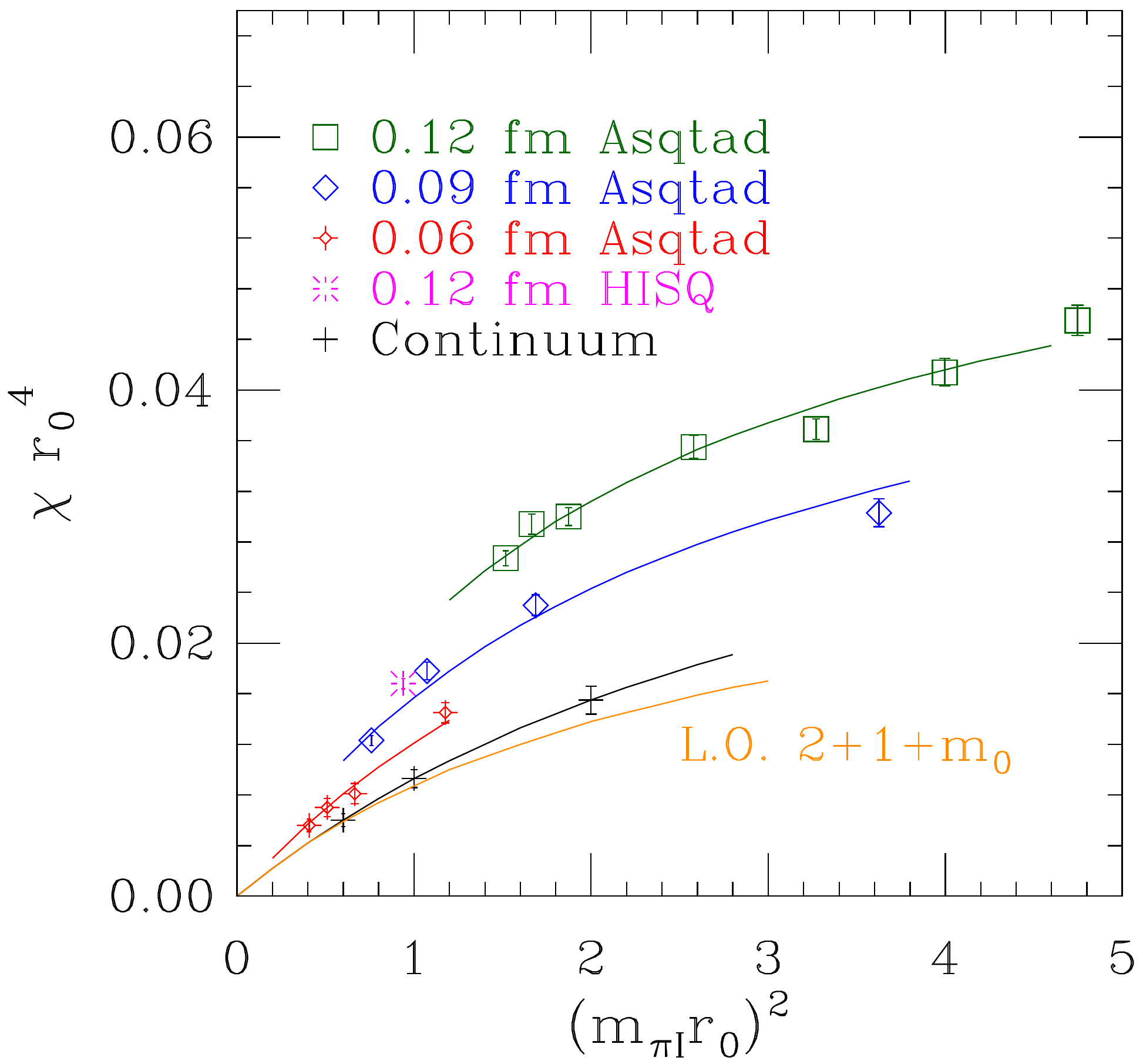}
\caption{\label{fig:asq top sus}Gluon topological charge susceptibility (MILC Collaboration). $r_0$ is the Sommer parameter.}
\end{minipage}\hspace{2pc}%
\begin{minipage}{18pc}
\includegraphics[width=18pc]{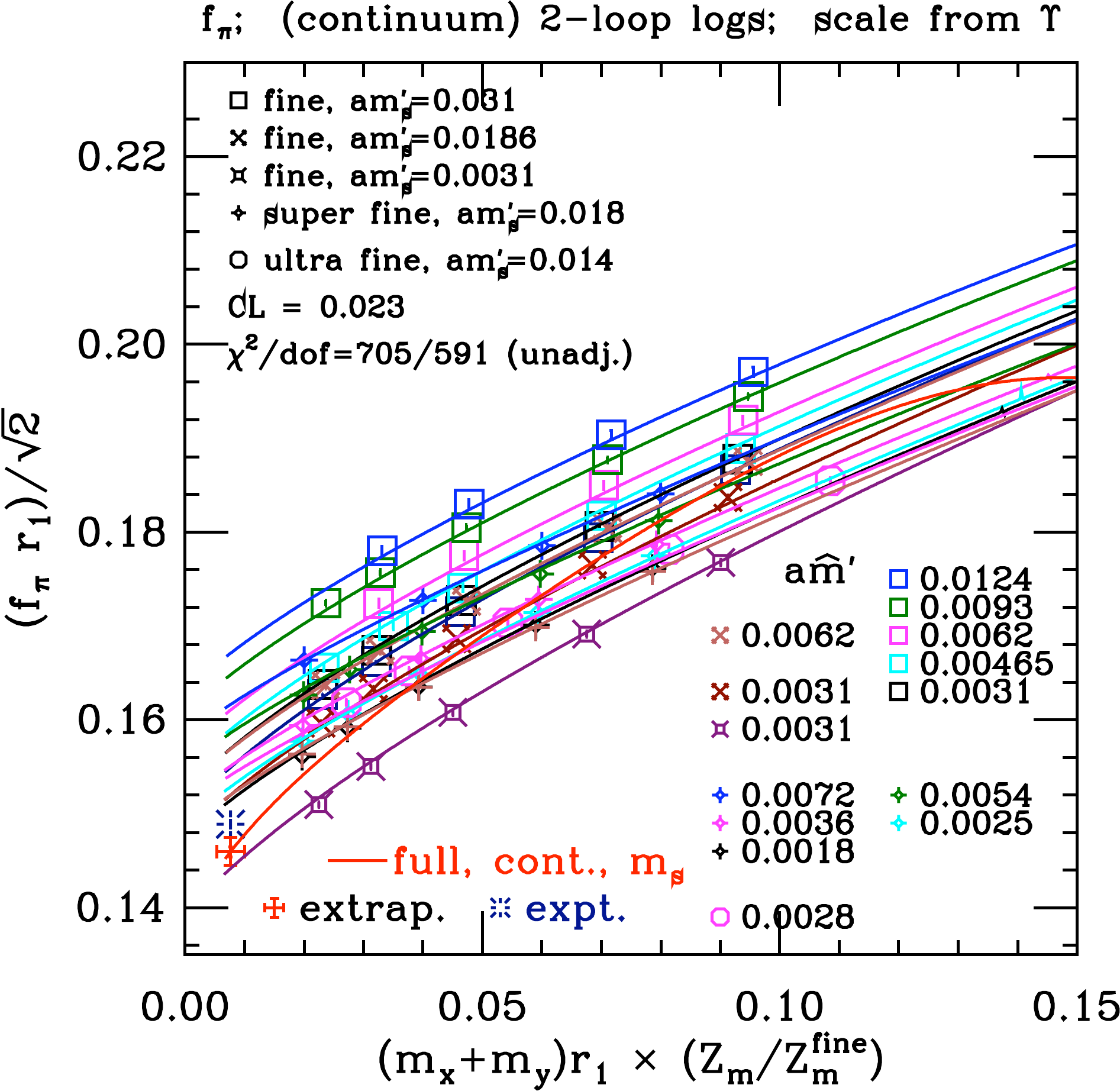}
\caption{\label{fig:asq decay}Pion decay constant (MILC Collaboration).}
\end{minipage} 
\end{figure}

The MILC Asqtad configurations are being used by several groups within USQCD to compute fundamental properties of QCD and the SM. The gluon topological charge susceptibility from MILC is shown in Fig.~\ref{fig:asq top sus}. When plotted with the non-Goldstone pion mass $m_{\pi1}$, it tends to vanish in the combined chiral and continuum limits, as it should. The pion leptonic decay constant is shown in Fig.~\ref{fig:asq decay}, accompanied by a multi-parameter fit of the data to chiral perturbation theory. The data are fit simultaneously with corresponding pion mass measurements and extrapolated in quark mass to the point where the pion mass agrees with its value from experiment. This fixes the value of the light quark mass at the ``physical point". Not only does this allow all other calculated quantities to be evaluated at the physical point, but the quark mass determined this way, once renormalized, is a fundamental parameter of the SM. 

A huge experimental effort is currently devoted to the study of
the decays and mixings of mesons with one heavy and one light
quark with the ambitious aim of observing phenomena beyond the SM.
The experimental measurements must be complemented, however, by precise
lattice QCD calculations of the corresponding weak matrix elements.

\begin{figure}[h]
\begin{minipage}{20pc}
\hskip 0pc
\includegraphics[width=20pc]{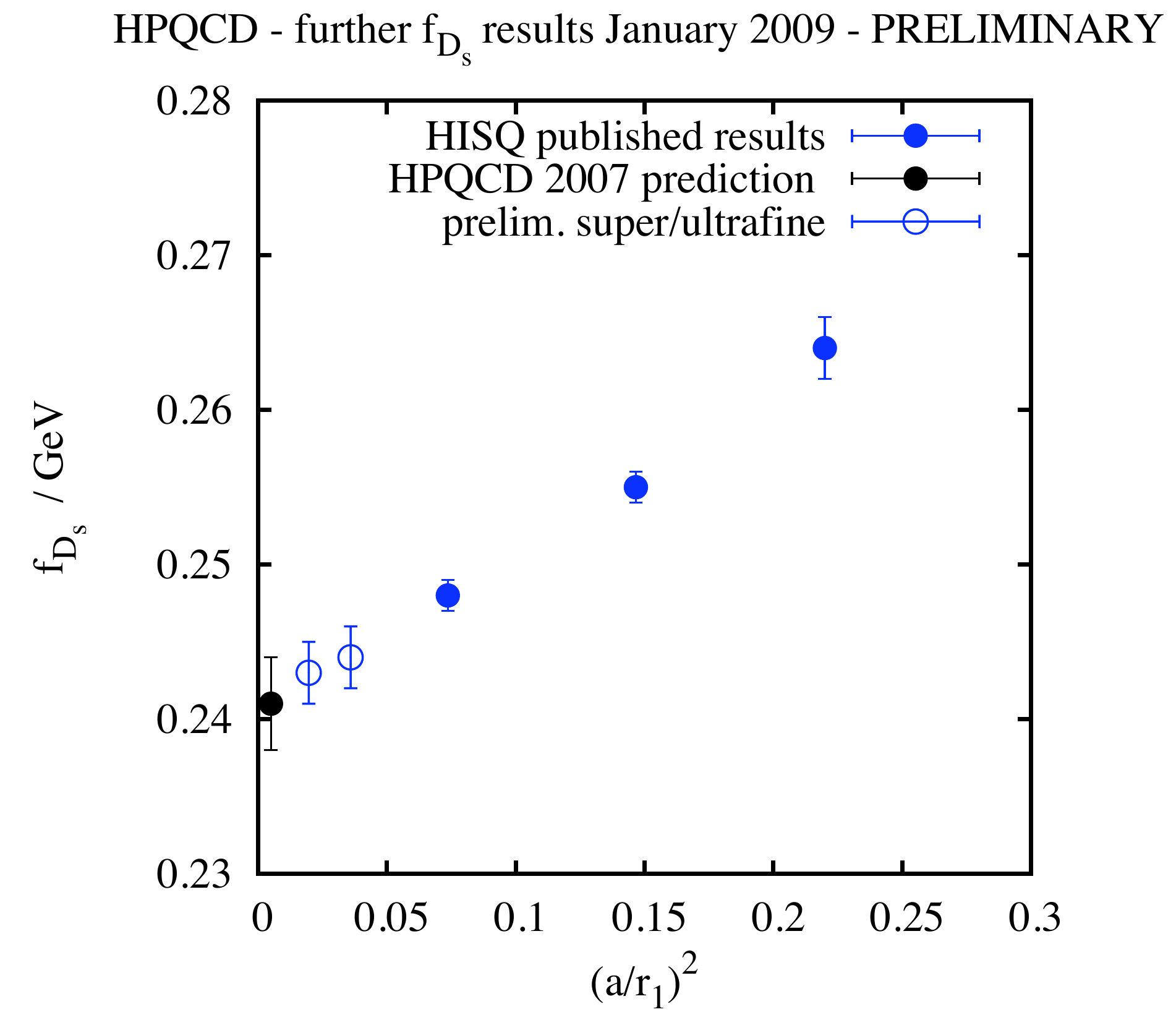}
\caption{\label{fig:c decay}$D_s$ meson decay constant (HPQCD Collaboration). }
\end{minipage}\hspace{2pc}%
\hskip -1pc\begin{minipage}{17pc}
\includegraphics[width=17pc]{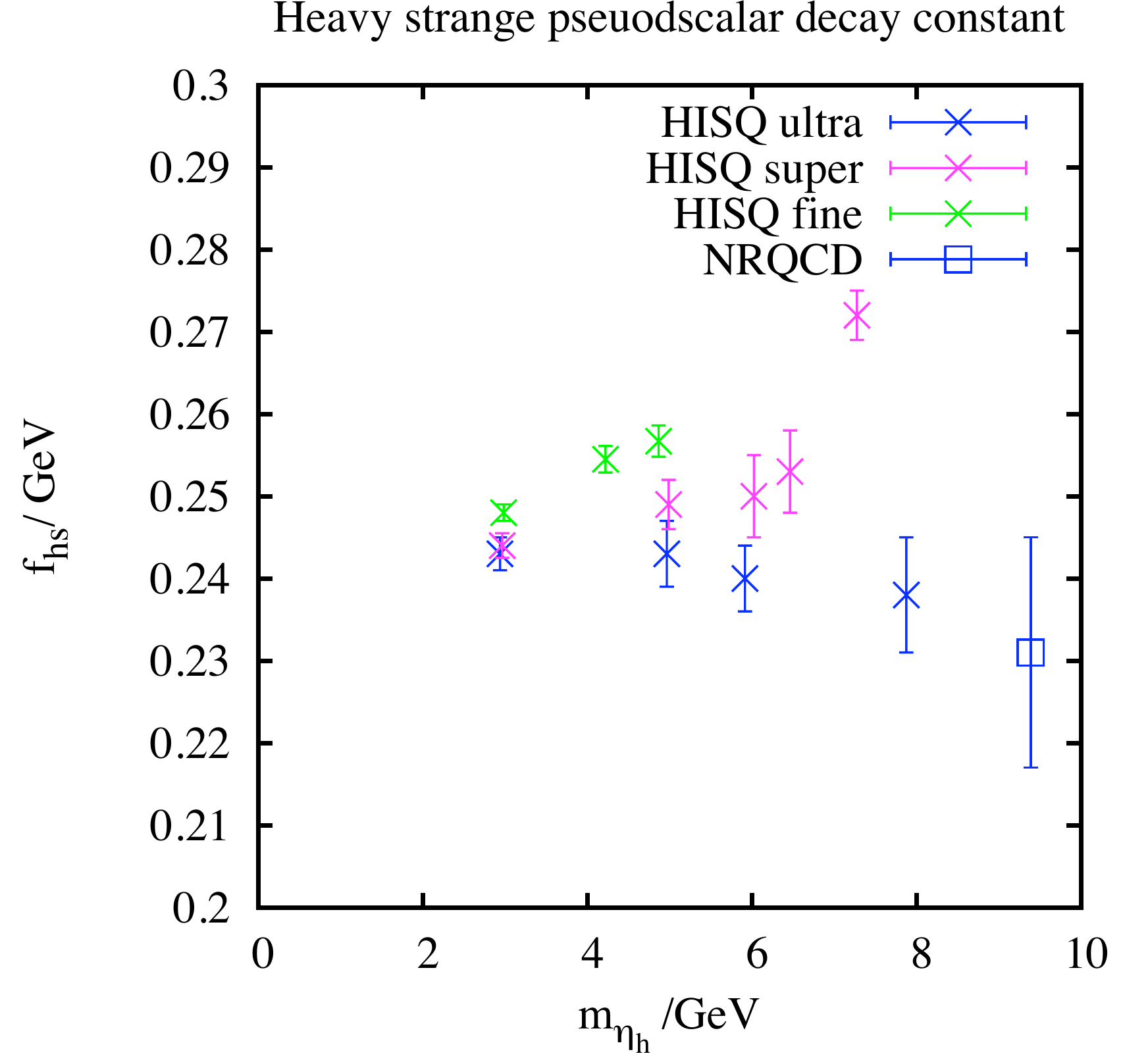}
\caption{\label{fig:b quark} Heavy-strange meson decay constant. Relativistic (HISQ) heavy quark computed with several lattice spacings (HPQCD Collaboration).}
\end{minipage} 
\end{figure}

The HPQCD calculation of the $D_s$ meson decay constant is shown in Fig.~\ref{fig:c decay}.
The new calculations on the super and ultra fine lattices are quite consistent with their earlier prediction which is 2.3 standard deviations lower than the 2009 measurement from the CLEO experiment.

The Fermilab/MILC group has been calculating heavy/light matrix elements on the Asqtad lattices, using Asqtad light valence quarks and clover improved Fermilab 
heavy valence quarks. They recently calculated the B meson decays 
$\bar B\to D^{*} l\nu_l$~\cite{Bernard:2008dn} and
$B\to \pi l\nu_l$~\cite{Bailey:2008wp} to determine the Cabibbo-Kobayashi-Maskawa (CKM) matrix elements $V_{cb}$ and $V_{ub}$, important SM parameters in the search for new physics.
These matrix elements will be calculated on the ultra fine ensemble, reducing the uncertainties
due to the heavy quark discretization, the largest  errors in this set of
calculations, by a factor of two.

With the advent of Leadership Class computers which can handle small $a$, large size lattices, HPQCD is investigating simulating a relativistic b quark ($m_b\sim 4$ GeV) directly on the lattice. Figure~\ref{fig:b quark} shows the result for the heavy-strange meson decay constant computed with highly improved staggered quarks (HISQ) for both the heavy and s valence quarks. As $a$ decreases, the decay constant trends toward the non-relativistic heavy quark (NRQCD) result. While inspiring, calculations at the b quark mass will require the next generation of computers.
HPQCD has also used HISQ valence quarks on the super fine and ultra fine lattices to compute the c quark mass.

\subsection{domain wall fermions}

The LHPC/RBC/UKQCD DWF gauge configurations listed in Table~\ref{tab:dwf confs}, along with a previously generated coarser ensemble with $a=0.114$ fm, are being used to study the light-strange K mesons (kaons) \cite{Allton:2008pn,Boyle:2007qe,Antonio:2007pb} and to investigate nucleon structure from QCD\cite{Syritsyn:2009np}. The former is an important compliment to the heavy-light meson investigation in the search for CP violation and new physics while the latter addresses an important aspect of QCD and is discussed in the talk on nuclear physics and lattice QCD by Edwards.

\begin{figure}[h]
\begin{minipage}{19pc}
\hskip 0pc 
\includegraphics[width=19pc]{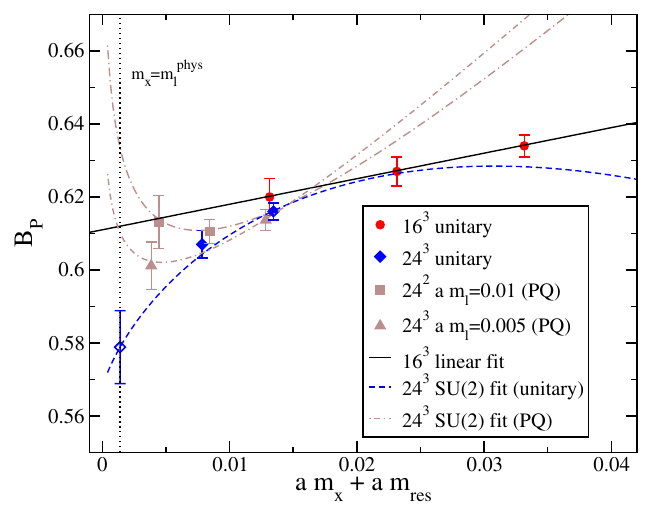}
\caption{\label{fig:bk} The kaon B parameter computed on the coarse DWF ensembles~\cite{Antonio:2007pb}. The (extrapolated) value at the physical point is shown at the lower left of the figure.}
\end{minipage}\hspace{2pc}%
\begin{minipage}{16pc}
\includegraphics[width=16pc]{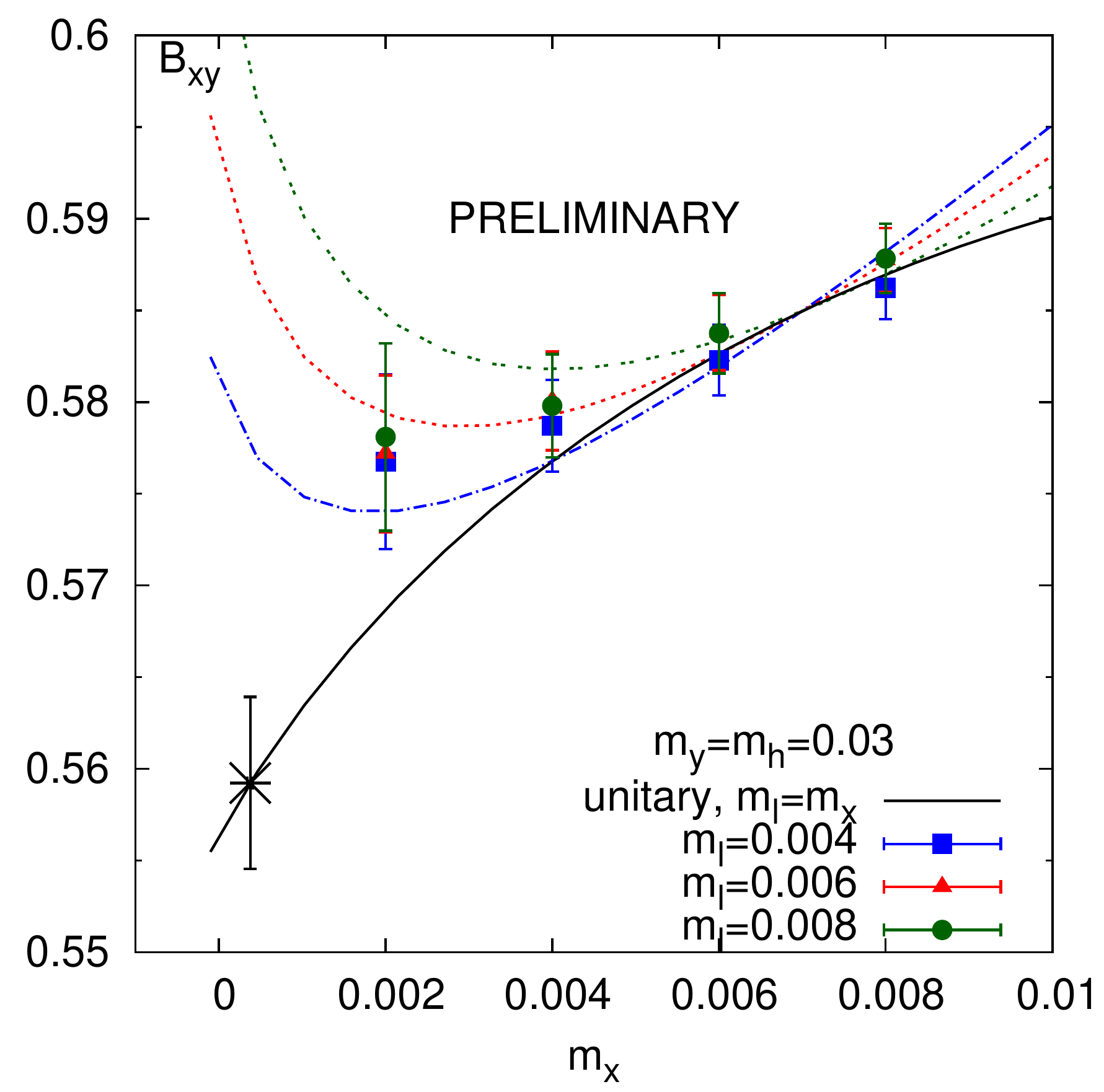}
\caption{\label{fig:bk fine}The kaon B parameter computed on the fine DWF ensembles (RBC and UKQCD collaborations).}
\end{minipage}
\end{figure}

Figure~\ref{fig:bk} depicts the most accurate calculation to date of the kaon B parameter $B_K$ which also provides a constraint on the vertex of the unitary triangle (geometrical representation of the CKM matrix) through the experimental measurement of K meson mixing since CP violation in the SM arises only from a single complex phase in the CKM matrix. The figure reveals how tricky it is to extrapolate to the physical point. Heavier light-quark points exhibit markedly linear behavior which belies the true behavior in the chiral limit. 
The current precision on $B_K$ is roughly 6 \% from all sources, but a large fraction is due to the lattice spacing error which is about 4\%. With the new fine DWF lattices, this error can be reduced significantly, constraining the SM even further.
Preliminary results for $B_K$ computed on the fine lattices are shown in Fig.~\ref{fig:bk fine}. The RBC and UKQCD collaborations are currently working to use both calculations to establish the continuum limit of $B_K$. Indicating its importance to the particle physics community, two other groups in USQCD are computing $B_K$ in the ``mixed-action" approach. Sharpe, {\it et al.}, use the Asqtad lattices with HYP-smeared staggered valence quarks~\cite{Bae:2008tb} while Laiho, {\it et al.}, use the Asqtad lattices but with DWF valence quarks~\cite{Aubin:2009jh}.

\begin{figure}[h]
\begin{minipage}{19pc}
\hskip 0pc
\includegraphics[width=19pc]{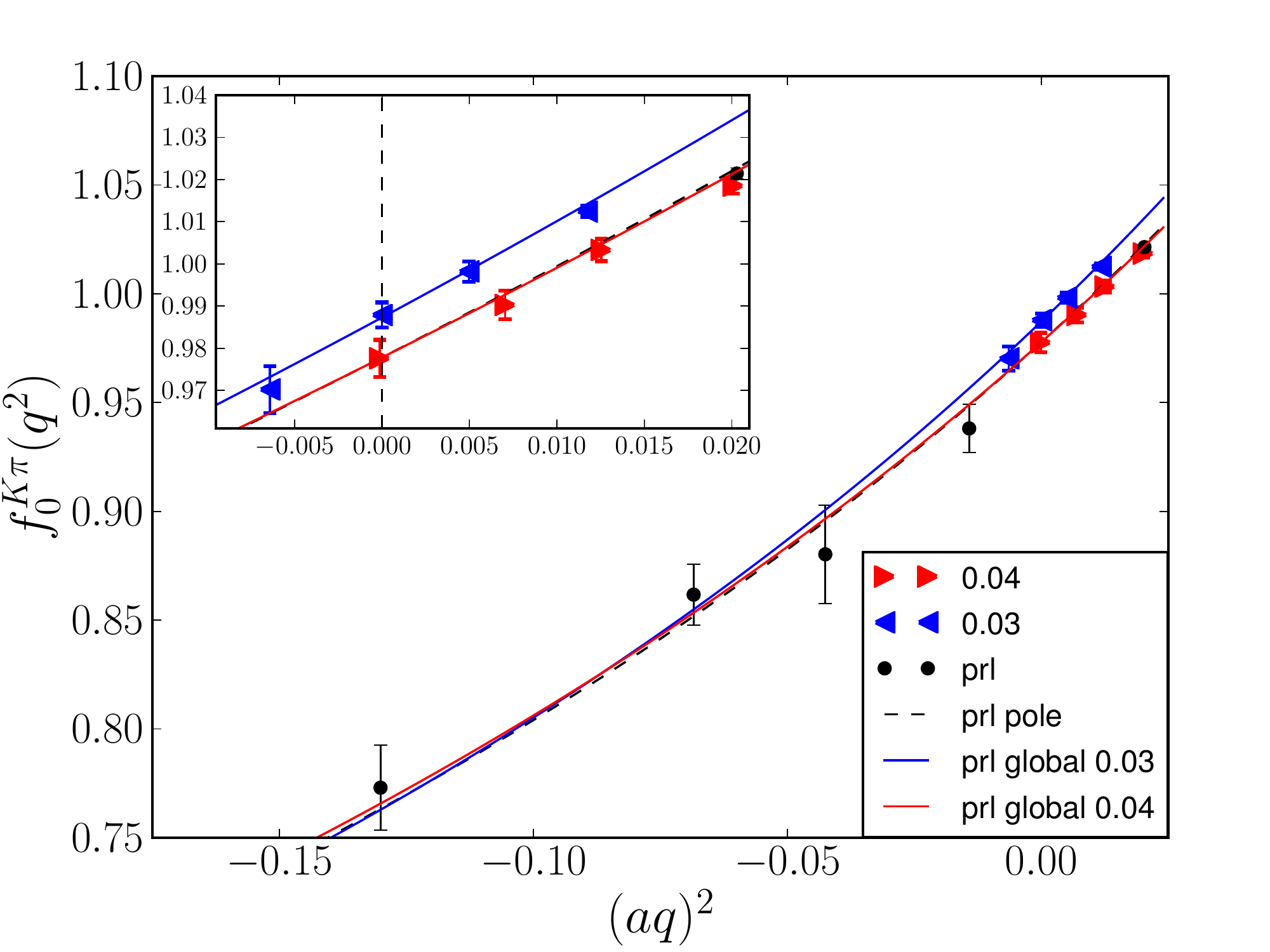}
\caption{\label{fig:kl3}Form factor $f_0(q^2)$ needed to extract $|V_{us}|$ from experimental measurement of the K meson decay $K\to\pi l \nu$, computed on the coarse ensemble\cite{Boyle:2007qe}.}
\end{minipage}\hspace{1pc}%
\hskip 0pc\begin{minipage}{18pc}
\includegraphics[width=18pc]{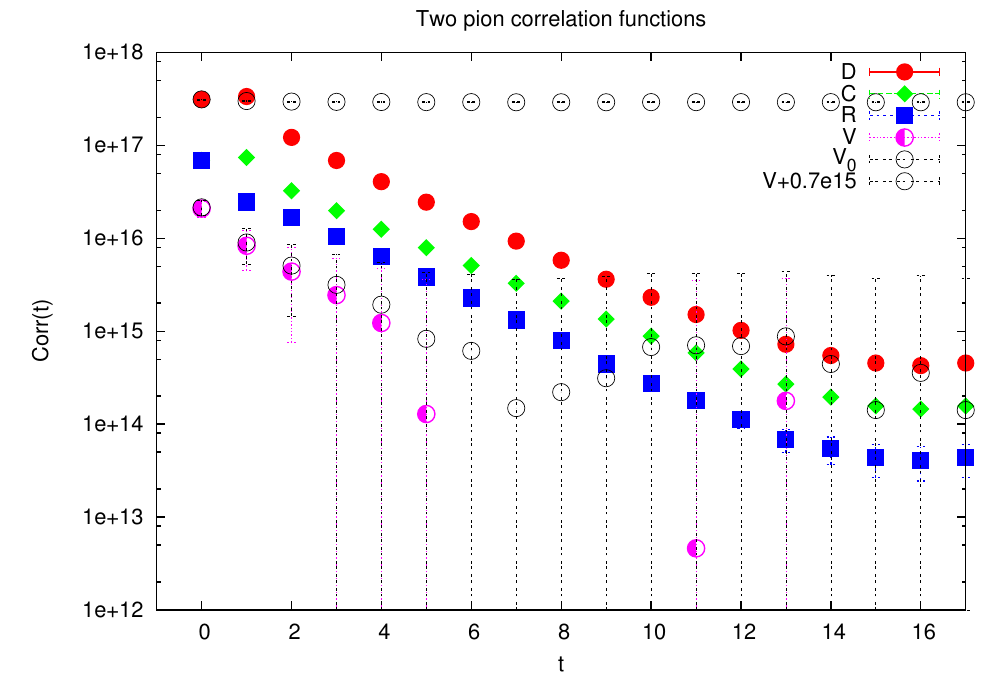}
\caption{\label{fig:pipi}Correlation function for $\pi\pi$ scattering. Different curves correspond to four distinct quark contractions in the correlation function (RBC and UKQCD collaborations).}
\end{minipage} 
\end{figure}

Another important precision test of the SM comes from unitarity of the CKM matrix, for example $|V_{ud}|^2+|V_{us}|^2+|V_{ub}|^2=1$ must hold; any deviation signals new physics. To extract $|V_{us}|$ from meson decays whose underlying quark decay process is $s\to u$, a non-perturbative form factor $f_{0}(0)$ must be determined. Figure~\ref{fig:kl3} shows $f_{0}(q^2)$ computed with 2+1 flavors of DWF\cite{Boyle:2007qe}.  The original calculation extrapolated to $q^2=0$ as required by the kinematics imposed by the lattice size and periodic boundary conditions. New points, computed using twisted boundary conditions and allowing simulations (nearly) at $q^2=0$, show excellent agreement.
As with $B_K$, analysis is underway using the fine lattices to take the continuum limit of $f_0(0)$.

Like the Asqtad program, a main goal of the DWF work is to investigate QCD near the chiral limit and confront experiment. 
On the coarse DWF ensemble, an important finding was that SU(3) chiral perturbation theory for 2+1 flavors of quarks does not work well when the s quark mass is near its physical value of roughly 100 MeV and only terms in the next-to-leading-order in the expansion are kept. It was much more accurate to use SU(2) chiral perturbation theory to extrapolate the light quarks to their physical point~\cite{Allton:2008pn}. In fact, this was done for the calculation of $B_K$ discussed above. Confirming this behavior on the fine lattices is underway. It also suggests running simulations at even smaller quark masses. We have started such a run where the pion mass $\sim 240$ MeV (last row in Table~\ref{tab:dwf confs}). However even with the prodigious power of Intrepid, the simulation is out of reach and must await the next generation of Leadership Class machines. In the meantime, one can back off of $a$, lower the quark mass, and still have a suitably large volume. The difficulty is that for DWF, increasing $a$ also increases the unwanted 
chiral symmetry breaking effects from finite 5th dimension. 

To tame the chiral symmetry breaking effects, we add an irrelevant term to the action, or equivalently, an auxiliary determinant (AuxDet for short) to the weight in the path integral~\cite{Vranas:1999rz,Fukaya:2006vs,Renfrew:2009wu}. The AuxDet allows for larger $a$ without the extra chiral symmetry breaking effects. DWF simulations with $m_\pi=180$ and 240 MeV are being run now at the ALCF. Besides strengthening the investigation of the chiral limit of QCD begun on the conventional DWF lattices, these large volume, small mass simulations open up an interesting new challenge: simulating decays of K mesons to two-pion final states. Important outcomes of such calculations are the direct CP violation parameter $\epsilon^\prime$ and the long standing $\Delta I=1/2$ rule for K decays. Figure~\ref{fig:pipi} shows preliminary results computed on the 16$^3$, coarse DWF ensemble for the isospin  0, two $\pi$ meson scattering correlation function. Such calculations are extremely demanding since they involve so-called ``disconnected" diagrams where the quark propagators are not connected  with each other. The disconnected diagrams require huge statistics to resolve. Hence, the $K\to\pi\pi$ decay project is to run on Intrepid in 2009 as the AuxDet lattices become available.

\section{Outlook}

Enabled by large machines like Intrepid, the gains achieved in lattice QCD simulations were not even dreamt of a few years ago.
As the continuum and chiral limits are approached, lattice QCD calculations are living up to the promise of solving the incredibly rich, complex, and beautiful theory of the strong interactions, a fundamental theory that is a cornerstone of particle and nuclear physics.
Still there is work to be done to reach these difficult limits, and we can't help but look to the next generation of computers. In the next five years, with sufficient but realistic resources ($O(100)$'s of Teraflop-years), QCD calculations may
be done at the physical point with several lattice spacings and sizes 
to take the continuum and infinite volume limits. 
The Large Hadron Collider (LHC) Era is set to begin.
Electro-weak symmetry breaking, supersymmetry, and beyond the SM phenomenology offer new challenges for lattice gauge theory.

\section{References}

\bibliographystyle{iopart-num}
\bibliography{proc-arXiv}

\vskip 15pt
\noindent{\bf Acknowledgements.}
I thank Bob Mawhinney for help with the discussion on tuning the RHMC algorithm. I also thank Peter Boyle, Norman Christ, Carleton Detar, Steve Gottlieb, Chulwoo Jung, Andreas Kronfeld, Junko Shigemitsu, Enno Scholz, and Bob Sugar for preparing various results covered in these proceedings.

\end{document}